\documentclass[jap,aip,reprint,longbibliography,amsmath,amssymb,twocolumn,superscriptaddress]{revtex4-1}
\usepackage{graphicx}
\usepackage{sidecap}
\usepackage{bm}
\usepackage{amsmath}
\usepackage{indentfirst}
\usepackage{times}
\usepackage{verbatim}
\usepackage{float}
\usepackage{color}
\usepackage{xcolor}

\begin{document}

\title{Understanding Magnetic Properties of Actinide-Based Compounds from Machine Learning }

\author{Ayana Ghosh}
\email{ayana.ghosh@uconn.edu}
\affiliation{Department of Materials Science $\&$ Engineering and Institute of Materials Science, University of Connecticut, Storrs, CT, 06269 - USA}
\affiliation{Theoretical Division, Los Alamos National Laboratory, Los Alamos, NM, 87545 - USA}
\author{Filip Ronning}
\affiliation{Institute of Materials Science, Los Alamos National Laboratory, Los Alamos, NM, 87545 - USA}
\affiliation{Condensed Matter and Magnet Science, Los Alamos National Laboratory, Los Alamos, NM, 87545 - USA }
\author{Serge Nakhmanson}
\affiliation{Department of Materials Science $\&$ Engineering and Institute of Materials Science, University of Connecticut, Storrs, CT, 06269 - USA}
\affiliation{Department of Physics, University of Connecticut, Storrs, CT 06269 - USA}
\author{Jian-Xin Zhu}
\email{jxzhu@lanl.gov} 
\affiliation{Theoretical Division, Los Alamos National Laboratory, Los Alamos, NM, 87545 - USA}
\affiliation{Center for Integrated Nanotechnologies, Los Alamos National Laboratory, Los Alamos, NM, 87545 - USA}

\date{\today}
(Report: LA-UR-19-25837)
\begin{abstract}
Actinide and lanthanide-based materials display exotic properties that originate from the presence of itinerant or localized \textit{f}-electrons and
include unconventional superconductivity and magnetism, hidden order; and heavy fermion behavior. 
Due to the strongly correlated nature of the 5\textit{f} electrons, magnetic properties of these compounds depend sensitively on 
applied magnetic field and pressure, as well as on chemical doping.
However, precise connection between the structure and magnetism in actinide-based materials is currently unclear. 
In this investigation, we established such structure-property links by assembling and mining two datasets that aggregate, respectively, 
the results of high-throughput DFT simulations and experimental measurements for the families of uranium- and neptunium-based binary compounds. 
Various regression algorithms were utilized to identify correlations 
among accessible attributes (features or descriptors) of the material systems and predict their cation magnetic moments and 
general forms of magnetic ordering. 
Descriptors representing compound structural parameters and cation \textit{f}-subshell occupation numbers were identified as most 
important for accurate predictions.
The best machine learning model developed employs the Random Forest Regression algorithm and can predict magnetic moment sizes and ordering forms
in actinide-based systems with 10--20\% of root mean square error.
\end{abstract}


\maketitle

\section{introduction}
Big-data driven approaches employing supervised, semi-supervised or unsupervised machine learning algorithms are becoming
tools of choice in materials physics, chemistry and engineering for the task of establishing yet unknown structure-property-performance 
relationships that may exist within a given family or class of materials.\cite{Rajan2015,Mueller2016,Butler2018,Graser2018,Rupp2015}
The success of these tools in elucidating hidden connections between the material or molecular structure and the resulting behavior
can be attributed to growing availability of databases collating theoretical and experimental materials data across disciplines.
In particular, databases aggregating the results of density functional theory (DFT) computations, which 
provide a reasonable compromise between high accuracy and computational costs and can also process fictitious materials structures,
are especially popular as components of prediction-driven strategies for materials design and discovery.\cite{Jain2013,Ong2015}  
A non-exhaustive list of examples demonstrating applications of machine learning algorithms in materials science
includes multiple investigations conducted for the families of technologically critical
(energy harvesting, storage and efficiency,\cite{Behler2011,Pozun2012,Snyder2012,Meredig2014,Schutt2014,Huan2015,Lookman2015,Seko2017,Stanev2018,Oliynyk2016,Mar2017} catalysis,\cite{Faber2015} photovoltaics,\cite{Pilania2016} etc.) 
and pharmaceutical (drug design,\cite{Ghosh2019,Lo2018} reaction mechanisms,\cite{Smith2018,Tropsha2015,Tropsha2016,Rupp2012} etc.) compounds. 
In addition to more generic traits originating from their general chemistry and radioactive behavior, lanthanide and actinide-based materials
exhibit a range of interesting properties associated with the filling of the 4\textit{f} and 5\textit{f} electron subshells.
In particular, the interplay of the hybridization of 5\textit{f} electrons with itinerant conduction electrons and the on-site Coulomb repulsions among those electrons is responsible for the behavior exhibited by actinides.
Such properties may include an emergence of magnetism\cite{Jaime2017} and colossal magnetoresistance at partial subshell fillings, as well as
unconventional superconductivity and magnetism,\cite{{Pfleiderer}} metal-insulator transitions,\cite{Seaborg} hidden magnetic order\cite{Kung2015} and the presence of heavy fermions.\cite{Coleman2007}
Due to strong correlation effects involving 5\textit{f}-electrons and their interactions with itinerant conduction electrons,
magnetic behavior of actinide-based systems is sensitive to applied pressure, magnetic field and chemical doping.
As a result, actinide-based materials are not only useful in nuclear applications but also constitute an interesting playground to push our fundamental understanding of correlated materials to the limit.
So far, only the 4\textit{f}-electron magnetism has been studied with DFT-based Machine Learning (ML) tools in a general context of ternary oxide compounds.\cite{Ceder2010}
%
A focused study of magnetic properties in 5\textit{f}-electron materials has not yet been reported. 
\begin{figure}
\centering
\includegraphics[width=1.05\columnwidth]{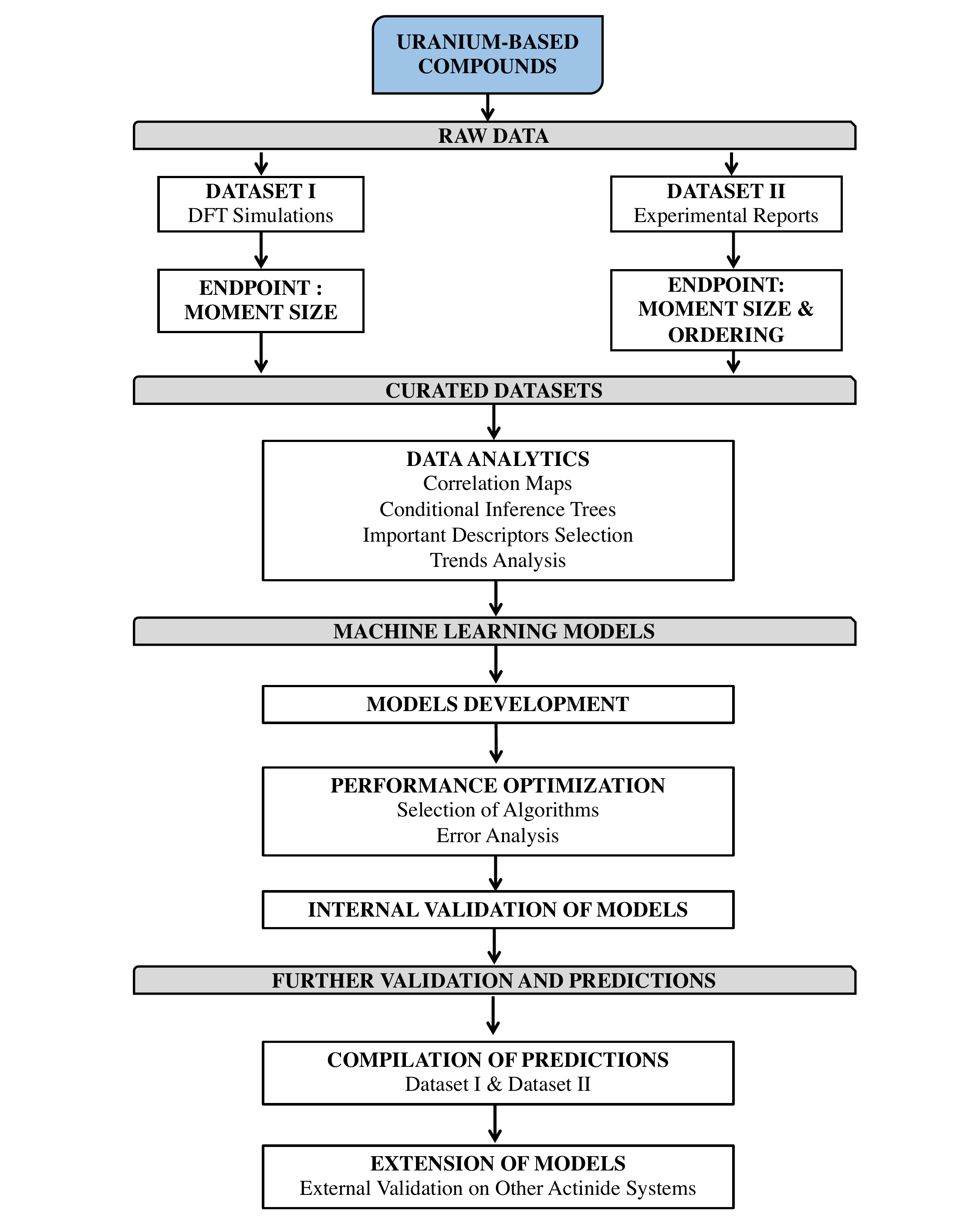}
\caption{
Flowchart outlining the main development stages involved in construction and validation of 
machine learning models for predicting cation magnetic moment size and magnetic ordering in actinide-based binary systems. 
Primary stages are shown as grey rectangles, while any necessary secondary stages are represented by white rectangles.
Some of the diagram elements introduced here are analyzed in further detail in Fig.~\ref{fig:struct_flow}(b) and accompanying text.
}
\label{fig:flowI}
\end{figure}
\begin{figure*}
\centering
\includegraphics[scale=0.5]{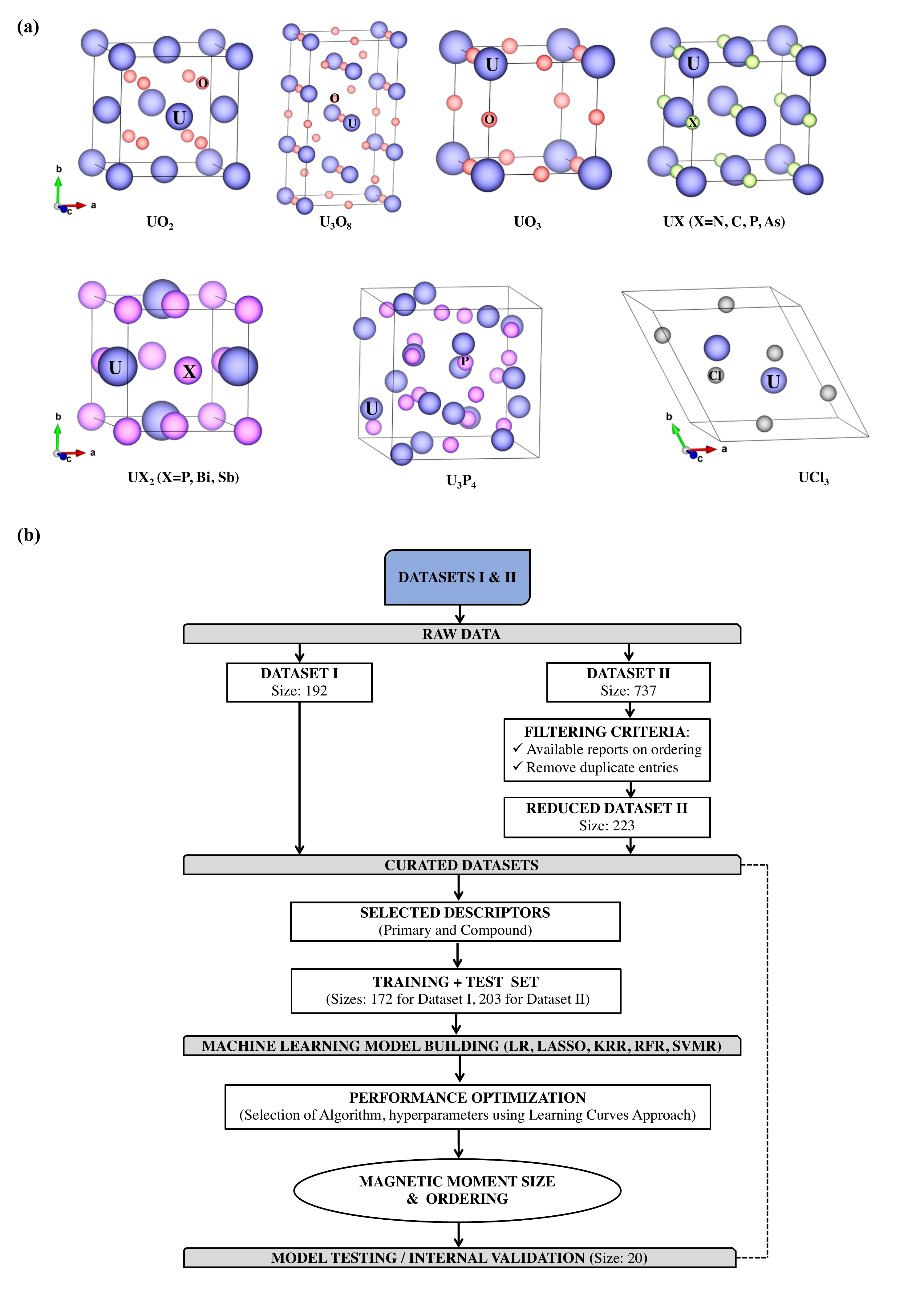}
\caption{(Color Online) (a) Structural models of 12 uranium-based binary compounds that were assumed in DFT computations for creating Dataset I. (b) Flowchart showing the stages of development of machine learning models to predict both moment size and ordering.}
\label{fig:struct_flow}
\end{figure*}

The main purpose of this work involves a systematic investigation of possible connections between the structure and magnetic properties 
for a variety of different actinide-based compounds in an attempt to
establish a general prescription for constructing families of machine learning models that
incorporate computational and experimental knowledge. 
Complementary utilization of data (See SM\cite{SM} for references)
originating from both of these sources is necessary for accurate assessment and prediction of the
magnetic properties of interest: average cation moment sizes (but not their ordering, which may be quite complex) can be easily 
extracted from DFT calculations, while magnetic ordering can be straightforwardly characterized
by experiments. 
Therefore, the task of compilation and curation of existing experimental reports on actinide-based systems is a critical part of this study.
The acquired data serves not only as a base for constructing machine learning models capable of predicting magnetic ordering,
but in some instances can also provide the necessary validation for the models utilizing only computational data. 
We must, however, point out the limitations of the experimental data, which naturally translates into the capabilities of machine learning models 
to make predictions: in most cases, only the major forms of magnetic ordering, such as paramagnetism (PM), ferromagnetism (FM) or antiferromagnetism (AFM) (\textit{as denoted by 1, 2 and 3, respectively in our work}), are reported, while information about the specific types of AFM or PM, or the orientation of magnetic moments with respect
to crystallographic axes is not given. 
The flowchart shown in Fig.~\ref{fig:flowI}, outlines the main stages of this study, which include compilation and curation of appropriate datasets, 
performing data analysis with standard data mining tools, construction of machine learning models and their following internal and external validation. 
The rest of the paper is organized as follows: Section II briefly reviews the standard methodology utilized for dataset acquisition and curation
as well as for the machine learning model building. 
Section III presents the results of statistical analysis for the contents of the both datasets 
followed by a discussion of predictive capabilities of the developed machine learning models  
in evaluating cation magnetic moment size and magnetic ordering in uranium- and neptunium-based compounds. 
Finally, some concluding remarks are provided in Section IV.

\section{Methods}
\subsection{Datasets}
First-principles calculations of average spin and orbital moments were performed using the projector augmented plane-wave (PAW) method
implemented in the Vienna Ab initio Simulation Package (VASP).\cite{Kresse1993,Kresse1996} 
The generalized gradient approximation (GGA) was adopted to represent the exchange and correlation interactions, 
with the GGA+$U_{\mathrm{eff}}$\cite{Dudarev1998} approach utilized to capture the strongly correlated nature of the 5\textit{f} electrons.
All computations were carried out with a 500 eV plane-wave cutoff energy using tetrahedron method with Blochl corrections with appropriate Monkhorst-Pack\cite{Monkhorst76} k-point meshes, which produced well converged results.
For the construction of Dataset I, which is built only on the data extracted from the DFT simulations,
the following twelve uranium-based binary compounds were utilized: 
UO$_2$, U$_3$O$_8$, UO$_3$, UN, UC, UP, UP$_2$, U$_3$P$_4$, UAs, UBi$_2$, USb$_2$ and UCl$_3$.
The magnetic structures of these compounds are well documented in the literature\cite{Troc66,Lebegue2006,Zhou2011,Wen2012} which is the primary motivation behind choosing them to build Dataset I. 
Geometrical structures for all of these compounds are shown in Fig.~\ref{fig:struct_flow}(a).
For each compound, initial lattice parameters and ionic positions
were obtained from Inorganic Crystal Structure Database (ICSD),\cite{ICSD}
after which eight individual variants were created by varying the Hubbard parameter $U_{\mathrm{eff}}$ between 
0 and 6 eV in 2 eV increments in the presence or absence of spin-orbit coupling. 
$U_{\mathrm{eff}}$ values from the same range have been used previously in a number of DFT-based investigations
of actinide compounds (See SM\cite{SM} for additional references).
%
Electronic and magnetic properties for each of the eight variants were evaluated both for the 
ICSD provided structural parameters and after optimization, which included relaxing the unit-cell shape and volume
to stresses below 0.1 kbar and all the ionic positions until the Hellman-Feynman forces below 10$^{-3}$ eV/\AA. 
Inclusion of data generated using structural parameters as obtained from both ICSD and DFT computations allows us to use information 
of varying fidelity\cite{Batra2019} levels which is helpful to improve accuracy in ML-based models.

For all computations to generate Dataset I, we consider antiferromagnetic ordering (AFM I configuration) for U cations with out-of-plane 
magnetic orientation along the $c$ axis. 
We restrict ourselves to studying only one type of magnetic configuration due to our interest in estimating moment sizes 
which do not vary significantly, if other configurations (e.g., FM) are selected instead. 
The choice of these initial AFM magnetic configurations for all 12 compounds is further discussed in 
SM\cite{SM} along with energy trade-offs between choosing in-plane vs. out-of-plane magnetic orientations. 
We report nominal differences ($<$0.3 $\mu_\mathrm{B}$) between these two AFM configurations as compared to 
average spin (1.64 $\mu_\mathrm{B}$) and orbital (2.82 $\mu_\mathrm{B}$) moment sizes for all of these compounds in the list,
which supports our approach in building Dataset I. 
Overall, Dataset I is built solely using DFT-simulations and comprises 16 cases for every compound, giving rise to a total of 192 entries. 
%

%
Dataset II was constructed by curating the results of 737 experimental reports of standard quality on uranium-based binary compounds, 
as found in the ICSD.\cite{ICSD}
Only structures that are stable at low temperature were considered, while data on any metastable high-temperature configurations
(that cannot be straightforwardly characterized by DFT calculations) was discarded.
After the removal of duplicate entries, 223 data points including information on magnetic properties (cation moment size and ordering) 
were obtained. 
These compounds were also categorized into three numbered groups according to the nature of the reported magnetic ordering in them: 
paramagnetic (compounds with local magnetic moment but no long range order present) group 1, ferromagnetic (compounds with magnetic spins aligned in the same directions) group 2, and antiferromagnetic (A-type or G-type) group 3.\\

The sizes of both datasets before and after curation, as well as the filtering criteria for Dataset II are shown in the top part of
the machine learning model development flowchart presented in Fig.~\ref{fig:struct_flow}(b).
For each dataset, 20 entries are kept aside for internal validations and the rest (172 for Dataset I and 203 for Dataset II ) are used for training (95$\%$) and testing (5$\%$) of ML-based models.

\begin{figure*}
\centering
\includegraphics[scale=0.5]{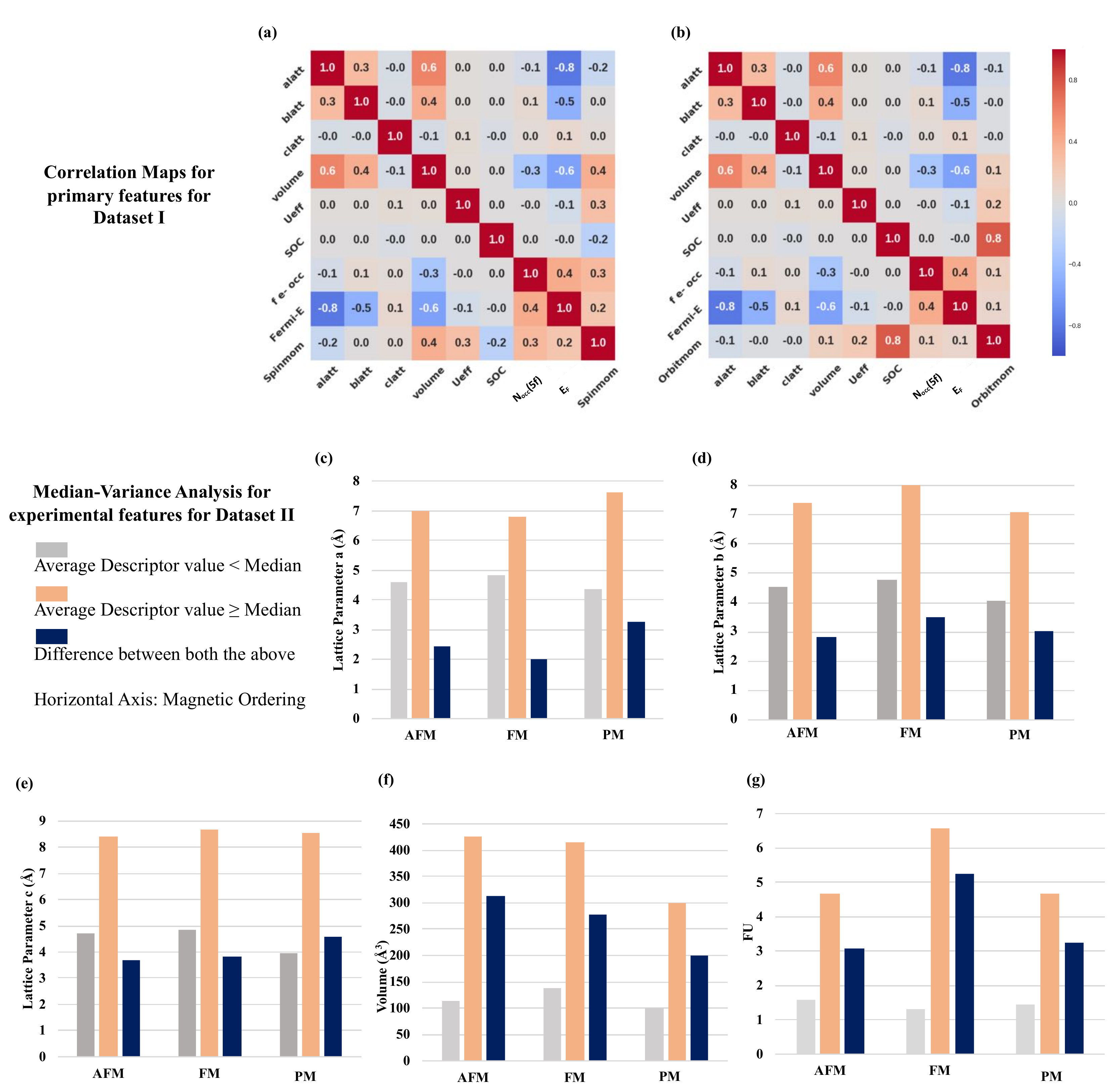}
\caption{
(Color Online) Correlation matrices representing Pearson correlation coefficients for primary descriptors applicable to Dataset I 
for (a) spin moment size and (b) orbital moment size endpoints.
The primary descriptor space for Dataset I consists of eight features: lattice parameters \textit{alatt}, \textit{blatt} and \textit{clatt}, and 
\textit{volume}, Hubbard parameter $U_{\mathrm{eff}}$, spin-orbit coupling strength {SOC}, 
cation 5\textit{f}-subshell occupation number \textit{$N_\mathrm{occ}(5f)$} and system Fermi energy level \textit{$E_F$}. 
(c-g) Dataset II is divided into two sets: below median value (grey), above median value (orange), difference (blue) for all experimental descriptors. 
The considered features are (c-e) lattice parameters (\AA) in all directions, (f) cell volume (\AA$^3$), (g) formula units (FU). 
The grouping is done based on the respective medians and type of magnetic ordering reported.
The median values for these features are (c) 5.588 \AA, (d) 5.5 \AA, (e) 5.62 \AA, (f) 191.85 \AA$^3$ and (g) 4.
}
\label{fig:corr_maps_med}
\end{figure*}

\subsection{Descriptors}
For Dataset I, the following eight primary descriptors were considered:
lattice parameters 
$(\frac{\text{magnetic unit cell parameters}} {\sqrt[3]{\text{number of actinide elements}}})$ \textit{alatt}, \textit{blatt} and \textit{clatt} (\AA),
atomic volume $(\frac{\text{magnetic unit cell lattice parameters}} {{\text{number of actinide elements}}})$ \textit{volume} (\AA$^3$), Hubbard parameter $U_{\mathrm{eff}}$ (eV), spin-orbit coupling strength SOC (eV), 
cation 5\textit{f}-subshell occupation number (number of total valence electrons of actinide element -- valence of anion) \textit{$N_\mathrm{occ}(5f)$} and system Fermi energy level \textit{$E_F$} (eV). 
For each primary descriptor $x$, additional compound descriptors were generated using 10 prototypical functions, namely, 
$x^2$, $x^3$, $\exp(x)$, $\sin(x)$, $\cos(x)$, $\tan(x)$, $\sinh(x)$, $\cosh(x)$, $\tanh(x)$, and $\ln(x)$, to allow for possible non-linearities in
the connections between the descriptor and endpoint properties. 
The descriptor space (exp. descriptors) for Dataset II contains all structural parameters (as defined before), 
number of formula units and \textit{$N_\mathrm{occ}(5f)$} , 
as extracted from the respective experimental reports.
Furthermore, for every entry in Dataset II, a matrix representation called Orbital field matrix (OFM)\cite{Pham2017,Pham2018} as implemented in 
a Python library\cite{matminer} was computed using distances between coordinating atoms, valence shells and Voronoi Polyhedra weights, providing information 
on the chemical environment of each atom in the unit cell. 
The OFM elements are defined in literature\cite{Pham2017,Pham2018} as following:
\begin{equation}
X^{\prime p}_{ij}=\displaystyle\sum_{k=1}^{n_p} o_i^p o_j^k\frac{\theta_k^p}{\theta_{\text{max}}^p}\zeta(r_{pk}),
\end{equation}
where, $i,j \in D = (s^1, s^2, \ldots , f^{14})$, for central $i$ and coordinating $j$ orbitals, respectively;  $o_i^p$ and $o_j^k$ are elements 
of one-hot vectors ($i,j$) of the electronic configuration $p$ and neighboring atom as indexed by $k$.
The weight of the atom $k$ in the coordination of the central atom at site $p$ is given by $\theta_k^p/\theta_{\text{max}}^p$, 
where ${\theta_k^p}$ is the solid angle determined using Voronoi polyhedron.
The number of nearest-neighbor atoms surrounding atom site $p$ is $n_p$.
The size and distance-dependent weight function is also included by $\zeta(r_{pk}$).
Two sets of ML-based models were constructed to predict magnetic ordering using experimental descriptors and OFM representation
for which the results are discussed later in Section III.

\subsection{Data Analytics}
Prior to the development of machine learning models, both datasets were analyzed using some standard data analytics techniques 
to determine the existence of any relationships between the structural descriptors and the targeted magnetic properties. 
The resulting inferences drawn from these analyses are complementary in choosing appropriate descriptors space and algorithms in later stages to develop ML-based models.  
The primary descriptor sets for Dataset I were subjected to a Pearson correlation filter to remove features that exhibit a 
high correlation with the other descriptors in each set. 
The same approach was also applied to the combined set of primary and compound descriptors as further discussed in SM.\cite{SM}
Additionally, for Dataset I we have employed a conditional inference procedure with Bonferroni-corrected significance 
(p-value $<$ 0.05) value, used as the splitting criteria (stopping rules) for each node while constructing trees as implemented in R version 3.4.2 via CTree algorithm.\cite{Rcodes} 
The splitting process is continued recursively throughout the whole Dataset I.
The results of this analysis are presented in the SM.\cite{SM}
We have utilized all of 223 entries in Dataset II to perform median analysis.
The dataset was divided into two subsets (below and above the median value) based on the median value for each descriptor.
For every descriptor, the magnetic ordering was assigned to the respective entries belonging to two resultant subsets. 
The bar charts (Fig.~\ref{fig:corr_maps_med} (c-g)) also report the differences between the mean values of the descriptors.
Larger difference for a particular descriptor should lead to a bigger variance when used in decision-tree type algorithms (such as Random Forest Regression). 

\subsection{Algorithms}
Five algorithms, including Linear Regression (\textit{LR}), Least Absolute Shrinkage and Selection Operator (\textit{LASSO}), 
Kernel Ridge Regression (\textit{KRR}), Random Forest Regression (\textit{RFR}) and Support Vector Machine Regression (\textit{SVMR}), 
as implemented in R\cite{Rcodes} version 3.4.2 were employed to construct machine-learning models from Datasets I and II. 
Regression-based algorithms were chosen over classification-based ones, as both endpoints of interest (magnetic moment size and ordering) 
were either computed for or assigned numerically to each entry in respective datasets.
Optimized hyperparameters used for all five algorithms are presented in the SM.\cite{SM}

\subsection{ML-based Model development and validation}
One of the main concerns of conventional machine learning based models development is the selection of data-points used in each of these steps. 
For both sets of models, the number of entries in each dataset is restricted to less than 230 entries, which may limit the accuracy of the models to be built.
The standard deviations in the predicted values for the test sets may be as large as 5-10\%, due to which any comparisons originating from a single set of predictions may be misleading.
This has been further explained in the SM,\cite{SM} where two such cases are compared.
In order to avoid any statistical bias, we have built a learning curve by varying training set size to estimate performance of all models. 
For each point plotted in a learning curve diagram,\cite{Ghosh2019} the average RMSE is reported as calculated using 1000 randomly generated (sampling with replacement) training and test set evaluations. 
The average RMSE for a training set size of $N$ is denoted $E_{\text{Train}}$(\textit{N}), whereas for the corresponding test set it is denoted $E_{\text{Test}}$(\textit{N}), however, in this case the size of the set is now the total number of points minus $N$. 
The test set RMSE gives the expected error for a given model whereas the difference between $E_{\text{Test}}$(\textit{N}) and $E_{\text{Train}}$(\textit{N}) is an estimation of how much variance or overfitting the model contains.
Both models predicting moment size and ordering built on Dataset I and II were tested using the \textit{internal} validation set, i.e. 20 entries (Fig.~\ref{fig:struct_flow}(b)) in each case that are kept aside on the model development stage as mentioned earlier.
Moreover, we applied these models to three \textit{external} sets of actinide-based binary and ternary compounds, for which information on either ordered moments or ordering is scarcely available in literature.
Validation on a dataset composed of materials that are dissimilar to that used in model development stage helps test the robustness and applicability of our conventional ML models. 
We note that the descriptor SOC is the strength of such coupling present in these systems as determined by the DFT based computations for Dataset I.
Hence, this is unavailable for all compounds present in the validation set.
As an alternative way to include this interaction in our models as applicable to the validation sets, we consider SOC by the form of its presence (1) or absence (0) instead of including its strength value. 
The $U_{\mathrm{eff}}$ can also be varied manually in the descriptor space depending on the estimate of this parameter to capture the strong correlation effect among the \textit{f}-orbital electrons.
In addition, the valence electrons of other atoms and OFM as applicable to the ternary set are also included in the feature space for testing models built on Dataset I and II. 
\begingroup
\begin{table*}
\centering
\caption{
Test set error $E_{\text{Test}}$(\textit{N}) and difference (overfitting) between the training $E_{\text{Train}}$(\textit{N}) and test set errors 
for the training set size $N = 50$ and the largest training test size $N_\mathrm{max}$, as obtained from the learning curves
constructed with five different regression algorithms.
The predicted endpoints are either size of spin or orbital magnetic moment in $\mu_\mathrm{B}$, or magnetic ordering.
Errors in prediction of magnetic ordering are with respect to the indices (1 - PM, 2 - FM, 3 - AFM) as introduced earlier in Section II.
}
\label{Magnetic_config}
\begin{ruledtabular}
\begin{tabular}{ccccccc}
Endpoint & Algorithm & $E_\mathrm{Test}(50)$ & $E_\mathrm{Test} - E_\mathrm{Train}(50)$ & $E_\mathrm{Test}(N_\mathrm{max})$ 
& $E_\mathrm{Test} - E_\mathrm{Train}(N_\mathrm{max})$ \\             
\hline
\hline
Spin moment & LR   & 0.61 & 0.44 & 0.33 & 0.12                    \\
& \textit{LASSO}   & 0.44 & 0.24 & 0.30 & 0.05                    \\
& \textit{KRR}  & 0.30 & 0.08 & 0.28 & 0.03                    \\
& \textit{SVMR}  & 0.84 & 0.04 & 0.83 & 0.01                   \\
& \textit{RFR} & 0.32 & 0.13 & 0.17 & 0.04                   \\ 
\hline
Orbital moment & LR &  1.46 & 0.88 & 0.96 & 0.24                    \\
& \textit{LASSO} & 0.90 & 0.14 & 0.80 & 0.10                   \\
& \textit{KRR}   & 1.07 & 0.4 & 0.96 & 0.21                    \\
& \textit{SVMR}   & 1.01 & 0.06 & 0.94 & 0.05                   \\
& \textit{RFR}  & 0.41 & 0.22 & 0.19 & 0.03                \\ 
\hline
Ordering & LR  & 1.77 & 1.14 & 1.02 & 0.50                   \\
& \textit{LASSO} &  1.32 & 0.82 & 0.98 & 0.41                    \\
& \textit{KRR} &  1.76 & 1.29 & 1.01 & 0.49                \\
& \textit{SVMR} & 0.86 & 0.03 & 0.85 & 0.02                   \\
& \textit{RFR} & 0.41 & 0.22 & 0.12 & 0.04                  \\ 
 \end{tabular}
\end{ruledtabular}
\label{table:errors}
\end{table*}
\endgroup

\begin{figure}
\centering
\includegraphics[width=0.9\columnwidth]{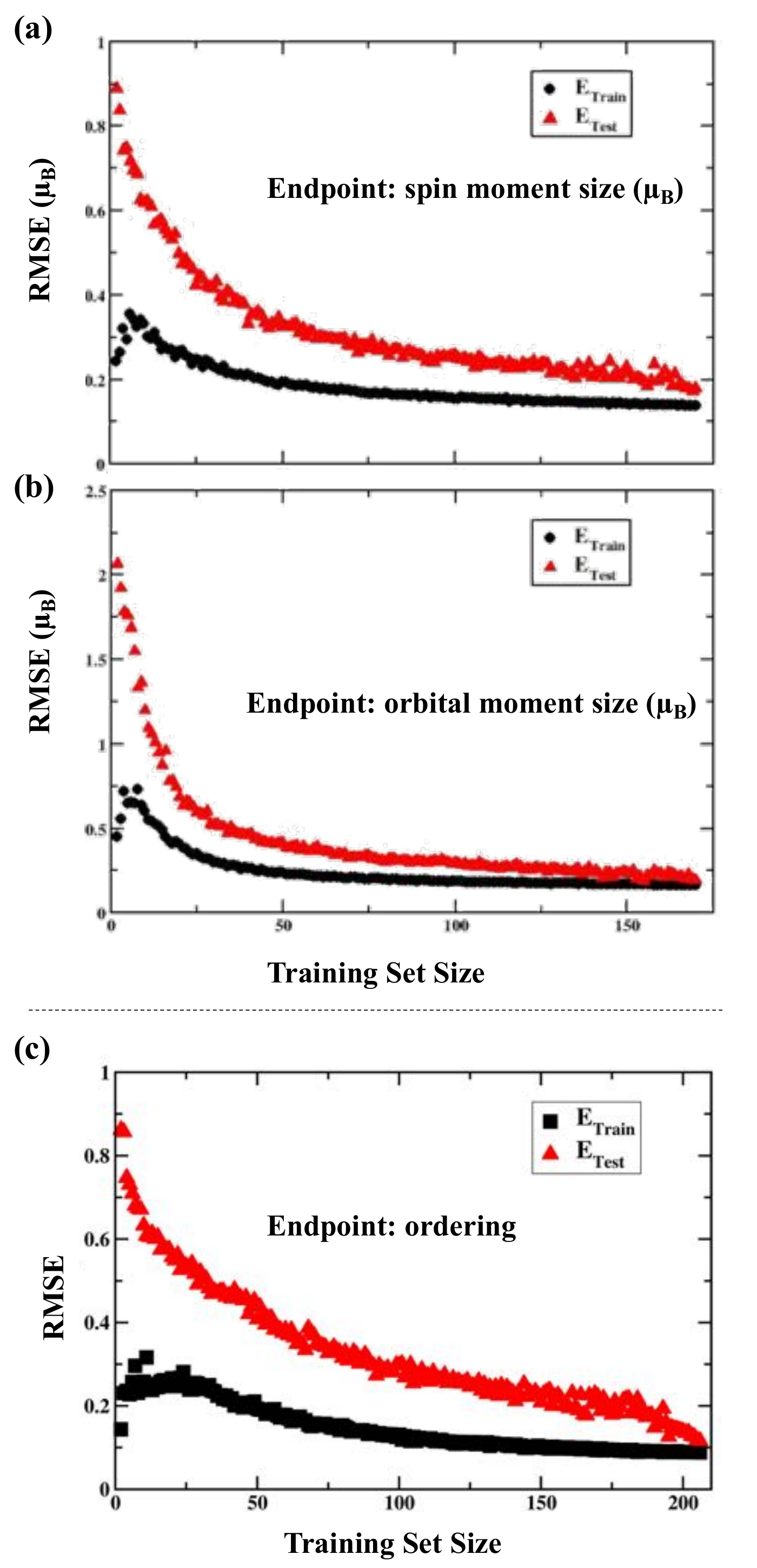}
\caption{
(Color Online) Learning curves for three machine learning models predicting (a) spin moment size, (b) orbital moment size and 
(c) magnetic ordering as constructed using the \textit{RFR} algorithm. 
$E_{\text{Train}}$ and $E_{\text{Test}}$ refer to average training and test set root mean square errors. 
Additionally, we report the average \% mean absolute test errors in predicting spin (a) and orbital (b) moment size are 14\%, 17\%. 
For ordering as represented in (c) this \% mean absolute test error is 8\%.
}
\label{fig:LR}
\end{figure}
\begin{figure*}
\centering
\includegraphics[scale=0.35]{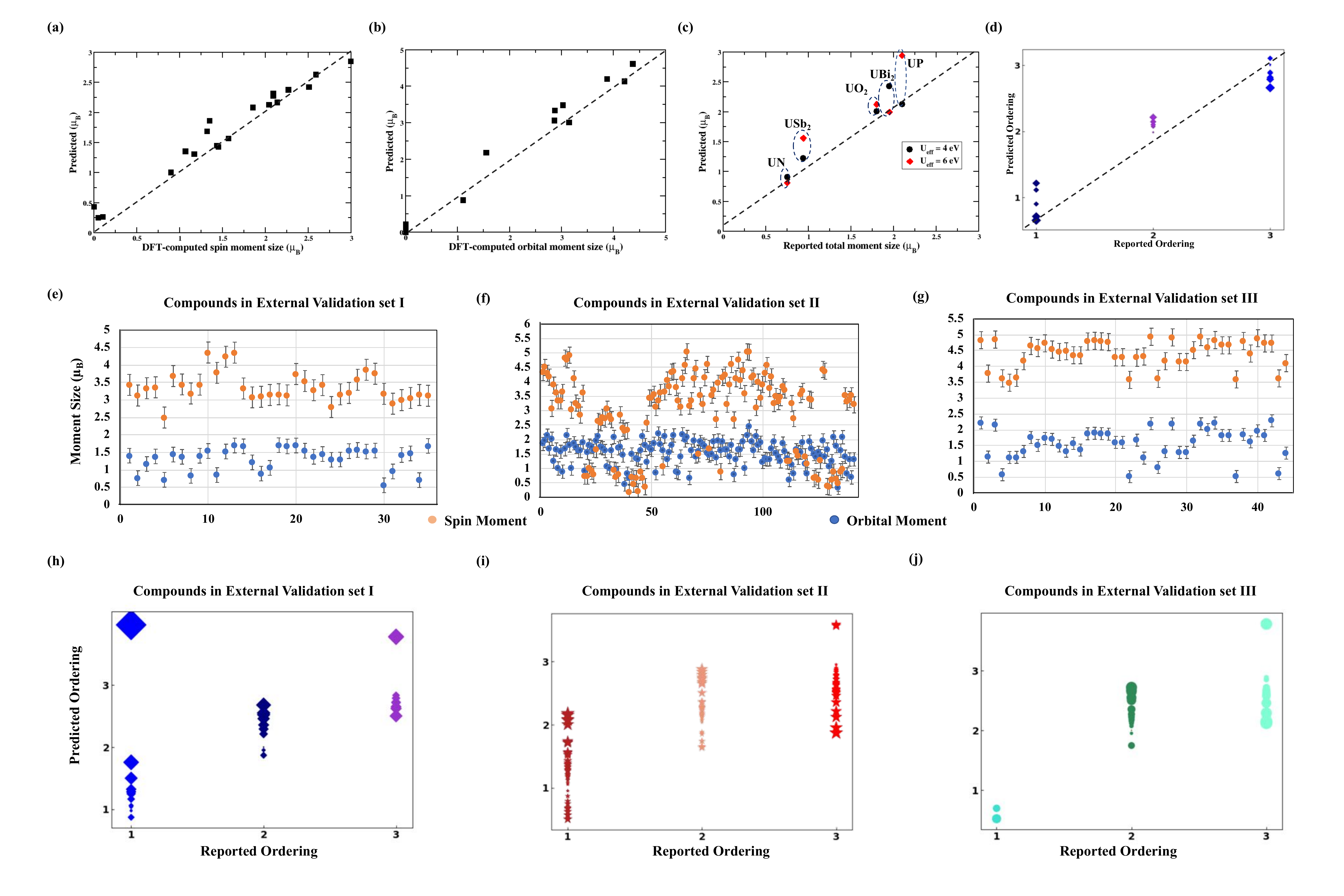}
\caption{
(Color online) ML-based models predictions of (a) spin and (b) orbital moment size on the \textit{internal} validation set from Dataset I. A comparison between  experimentally reported and predicted total moment size is represented in (c) for selected (for which moment sizes are available in literature) compounds in this set. Prediction on the \textit{internal} validation set from Dataset II is shown in (d). The radius of each marker represents the absolute error in each prediction for each entry present in the set as clustered together based on their ordering. (e-g) Independent predictions of spin and orbital moment size for all compounds present in three \textit{external} validation sets using two models solely built on Dataset I. The error bars in each case represent the average root mean square errors in these predictions as obtained by averaging over 1000 such ML-based models predictions. (h-j) Comparative predictions of magnetic ordering for all compounds listed in three \textit{external} validation sets using OFM-based model built on Dataset II.
The values of each prediction is obtained by averaging over 1000 such respective ML-models based predictions.
}
\label{fig:plot_validation}
\end{figure*}

\section{Results and Discussion}
Here, we discuss the results obtained from preliminary data analysis and ML-based models as applicable to both Dataset I, II, internal and external validation sets in sequential order, with each primary step highlighted throughout the section.
From the \textbf{correlation analysis} as presented in Fig.~\ref{fig:corr_maps_med}(a) on the set of primary features, it is evident that \textit{volume}, \textit{$N_\mathrm{occ}(5f)$} and \textit{$E_F$} have large Pearson coefficients with respect to the endpoint (spin moment size), suggesting that these features should be included in the descriptor space when building ML-based models on the later stage to predict spin moment size. 
However, since \textit{$N_\mathrm{occ}(5f)$} and \textit{$E_F$} are highly correlated to each other, we only keep \textit{$N_\mathrm{occ}(5f)$} as one of the primary descriptors.
We note that the \textit{$E_F$} as obtained from the DFT based computations is relative and can be placed anywhere in the energy band gap between the occupied and unoccupied states. 
Therefore the estimation of charges as accounted by the \textit{$N_\mathrm{occ}(5f)$} feature is more appropriate to retain information on the electron density of states which is important in predicting magnetic moment size. 
The SOC has the highest correlation coefficient with respect to the endpoint in the matrix, where the endpoint is orbital moment size as shown in Fig.~\ref{fig:corr_maps_med}(b). 
This is expected for the primary feature that accounts for strong SOC originating from 5\textit{f} shell electrons.
This approach as extended to space of primary and secondary features shown in the SM\cite{SM} has allowed us to shortlist 61 features by excluding highly-correlated ones (correlation factor $>$ 0.85). 
Next, the \textbf{conditional inference} procedure as employed on Dataset I and presented in the SM\cite{SM} shows that \textit{alatt}, \textit{volume}, \textit{U$_{\textit{eff}}$} and SOC are the top features capable of grouping the data well for spin and orbital moment size as endpoints, respectively.
We have used a slightly different approach (\textbf{median analysis}) to analyze Dataset II by determining medians and differences in average values of the features as explained in Section II.C.
Based on the median values of the five features, the Dataset II is divided into subgroups followed by assigning the corresponding ordering and calculating the difference (variance) in average feature values for the subgroups as shown in Fig.~\ref{fig:corr_maps_med}(c-g). 
The median values for each of these experimental descriptors (\textit{lattice parameters}, \textit{volume}, \textit{number of formula units}) are 5.588 \AA, 5.5 \AA, 5.62 \AA, 191.85 \AA$^3$ and 4 respectively.
There is total 112 entries in the subgroup where \textit{alatt} $>=$ median value of \textit{alatt}, out of which 24 are AFM, 44 are FM and 44 are PM.
For the similar subgroups formed by other features such as \textit{blatt}, \textit{clatt}, \textit{volume}, \textit{number of formula units}, the number of entries are listed as: (26 - AFM, 50 - FM, 36 - PM), (30 - AFM, 46 - FM, 36 - PM), (24 - AFM, 48 - FM, 40 - PM) and (42 - AFM, 70 - FM, 51 - PM), respectively.
%

%
Both conditional inference tree and median-variance methods suggest that the decision tree type of algorithm may have better performance compared to other regression algorithms, if used to build ML models for predicting moment size and ordering.
Overall, these investigations performed using standard data analytics techniques are useful for cultivating some advance knowledge about the available data and identification of features important to predicting endpoints as well as any inconsistencies present in the datasets. 

The primary results produced by the \textbf{ML-based models} are presented as the learning curves in Fig.~\ref{fig:LR}.
For all three cases, the \textit{RFR} algorithm outperforms the other algorithms significantly, as shown in both Fig.~\ref{fig:LR} and Table ~\ref{table:errors}. 
The average spin moment size of compounds used for training these models is 1.64 $\mu$$_{\text{B}}$.
For the orbital moment size, the average is 2.82 $\mu$$_{\text{B}}$, counting only the training data points computed by including the SOC. 
The total moment size can be obtained using the vector sum of both spin and orbital moments, pointed in the opposite direction to each other due to Hund's rule for \textit{f}-electron shell that is less than half filled.  
The average $E_{\text{Test}}$ values in predicting spin and orbital moment size are 0.17 $\mu$$_{\text{B}}$ and 0.19 $\mu$$_{\text{B}}$, respectively as mentioned in Table ~\ref{table:errors}.
In both cases, \textit{RFR} overfits the data by approximately 4\%. 
The LASSO, KRR and SVMR algorithms on the other hand show minimal overfitting but plateau at higher $E_{\text{Test}}$(\textit{N}) as shown by comparative learning curves (see SM\cite{SM} for more details).
The average RMSE in predicting magnetic ordering using Dataset II is 0.12.
The OFM representation plays a key role in significantly improving the model performance (RMSE reduced by $\sim$30\%) to predict magnetic ordering by including information on the valence shells and local coordination environment of the system. 
The details of the errors evaluation in the learning curves for both models are discussed in the SM.\cite{SM}
The most important features (structural parameters and \textit{f}-subshell occupation numbers) identified by \textit{RFR} algorithm based models comply with that found earlier by the data analytics techniques for both moment size and ordering. 

We have also compared the average nearest neighbor distances (see SM\cite{SM}) for every entry in Dataset II to the Hill limit\cite{Hill70} that provides restrictions, under which magnetic ordering occurs in actinide systems. 
Our results are in good agreement with the Hill limit for uranium ions (3.4\AA - 3.6\AA). 
This establishes a physical significance behind lattice parameters, which are identified as important features to predict both moment size and ordering.
The average lattice parameters obtained from Dataset I are comparable within 15\% to entries present in Dataset II that reportedly exhibit antiferromagnetic ordering at low temperatures.
This along with results from median-variance analysis performed on Dataset II also provides quantitative measure for structural parameters to observe a specific type of magnetic ordering (in high likelihood). 
For example, a uranium-based binary compound with \textit{alatt} $>=$ 5.58\AA,\linebreak
is more likely to exhibit antiferromagnetic ordering at low temperatures. 
Three of the models were tested first on the \textit{internal} \textbf{validation} sets kept aside in Dataset I and Dataset II.
We employed a comparable approach to the learning curve by reporting the average moment size and ordering for each entry as obtained by averaging over 1000 ML-based models predictions.
Figure~\ref{fig:plot_validation} (a and b) shows predictions made on the \textit{internal} validation set acquired using Dataset I. 
The average RMSE for the spin, orbital moment size predictions on the \textit{internal} validation sets (for $U_{\mathrm{eff}}$ = 4 eV) are 0.20 $\mu$$_{\text{B}}$, 0.25 $\mu$$_{\text{B}}$, respectively. 
For compounds such as UN, USb$_2$, UO$_2$, UBi$_2$ and UP, total moment sizes are available in the literature, and can be used to compare the ML-based models predictions for $U_{\mathrm{eff}}$ = 4 and 6 eV as represented in Fig.~\ref{fig:plot_validation}(c).
The average RMSE for the prediction of total moment size for these compounds are 0.32 $\mu$$_{\text{B}}$ and 0.35 $\mu$$_{\text{B}}$ for $U_{\mathrm{eff}}$ = 4 and 6 eV, respectively.
This analysis also signifies the dependence of moment size on $U_{\mathrm{eff}}$ as captured by the ML models.
Our ML-based regression algorithm can predict numerical values marking  AFM, FM and PM with an average RMSE of 0.15 as shown in Fig.~\ref{fig:plot_validation}(d), where the predictions are compared with that reported within the \textit{internal} validation set of Dataset II.
%
%
%

%
Finally, to assess the performance and applicability of these models to other actinide systems, we have compiled three \textit{external} validation sets, first two containing uranium-based binary and ternary compounds\cite{Sechovsky} and a third dataset consisting of neptunium-based binary compounds. 
The results of these predictions using the same averaging technique as applied before to the \textit{internal} validation sets are shown in Figs.~\ref{fig:plot_validation}(e-g) and ~\ref{fig:plot_validation}(h-j).
We note that most of the predictions (Fig.~\ref{fig:plot_validation} (e-g)) of the \textit{moment size} cannot be validated due to the scarcity of experimental information.
However, predictions of magnetic \textit{ordering} indices that classify compounds into three simple groups of AFM, FM and PM are comparable with those reported in the literature\cite{Sechovsky} without accounting for the exact spin textures. 
The \textit{external} Set I of binary compounds include 34 different uranium-based compounds U$_2$C$_3$, U$_3$As$_4$, U$_3$Bi$_4$, U$_3$Sb$_4$, UAl$_2$, UAl$_3$, UAs$_2$, UB$_2$, UB$_4$, UBi, UCo$_2$, UFe$_2$, UGa$_3$, UGe$_2$, UGe$_3$, UIn$_3$, UIr$_2$, UIr$_3$, UIr, UMn$_2$, UNi$_2$, UPb$_3$, UPd$_3$, UPt$_2$, UPt$_5$, UPt, URh$_3$, US, USb, USe, USi$_3$, USn$_3$, UTe and UTl$_3$. 
These are not present in either of the Datasets (I \& II) used in model development reported above.
The top 5 most common structure types in this list belong to cubic crystal family with space group numbers 221, 220, 225, 227 and 191.
There are 9 compounds that exhibit AFM, 10 with FM and the rest have PM ordering at low temperatures.
The average RMSE (based on ML-model built on Dataset I) in predicting moment size for these 9 compounds are 0.21 $\mu$$_{\text{B}}$ and 0.30 $\mu$$_{\text{B}}$ respectively for spin and orbital moment size.
For ordering, the average RMSE is 0.23 (based on ML-model built on Dataset II) as predicted for all 34 compounds, shown in Fig.~\ref{fig:plot_validation} (h). 
\textit{External} Set II containing ternary compounds has 141 different entries, out of which 51 exhibit AFM ordering, 31 show FM ordering and rest are paramagnetic at low temperatures.
These compounds commonly belong to families of orthorhombic, tetragonal, hexagonal and cubic crystal systems with the five most common space groups being 62, 139, 123, 127 and 189. 
We predict moment sizes (based on ML-model built on Dataset I) for these 51 compounds with average RMSE of 0.23 $\mu$$_{\text{B}}$ and 0.28 $\mu$$_{\text{B}}$ for spin and orbital moments respectively, whereas this error is 0.24 as depicted in Fig.~\ref{fig:plot_validation} (i) for estimating ordering (based on ML-model built on Dataset II).
\textit{External} Set III has a total of 44 entries of 35 unique neptunium-based compounds NpAl$_2$, NpAl$_3$, NpAl$_4$, NpGa$_2$, NpGa$_3$, NpIn$_3$, NpIr$_2$, Np$_2$N$_3$, NpNi$_2$, Np$_2$O$_5$, Np$_2$Se$_5$, Np$_3$S$_5$, Np$_3$Se$_5$ NpAs, NpAs$_2$, NpB$_2$, NpC, NpCo, NpFe$_2$, NpGe$_3$, NpIn$_3$, NpMn$_2$, NpN, NpN$_2$, NpNi$_2$, NpO$_2$, NpOs$_2$, NpP, NpPd$_3$, NpS, NpSb, NpSb$_2$, NpSi$_2$, NpSi$_3$ and NpSn$_3$.
The most common structure type among these compounds is rocksalt cubic followed by Laves phase cubic and Auricupride.
These three structure types are also common among compounds in Dataset I and II.
Hence, it suggests that the models may be capable of predicting the endpoints with similar accuracy as reported earlier in the section.
In \textit{external} Set III, 17 entries have AFM ordering as reported in literature. 
These 17 entries are used to validate the models (based on ML-model built on Dataset I) predicting spin and orbital moment size with average prediction RMSE of 0.19 ($\mu$$_{\text{B}}$) and 0.27 ($\mu$$_{\text{B}}$), respectively.
All 44 entries are considered to predict magnetic ordering using model built on Dataset II.
The average prediction error for ordering is reported as 0.14 for this set as shown in Fig.~\ref{fig:plot_validation} (j). 
Overall, the ML-based models built on Dataset I and II are capable of delivering reasonable predictions of moment size and ordering for actinide-based binary and ternary compounds.
\section{Conclusions}
In conclusion, we have compiled two datasets containing both computational and experimental reports on magnetic properties of uranium-based binary compounds. 
Through various data analytics techniques, we have identified several descriptors that are critical for understanding magnetism occurring in such systems, even before building any predictive models.
Later, these were used in developing sets of machine learning models to predict moment size and ordering.
We have also extended this approach to other actinides and assessed the performance of the models. 
Currently the models trained on Dataset I can only predict moment size for AFM ordered structures. 
Predicting magnetic ordering using estimation of nearest, next nearest neighbor exchange parameters require additional DFT computations for other magnetic configurations, which is beyond the scope of the current work. 
Overall, this general prescription employing both computational and experimental results to construct machine learning models describing magnetism in actinide-based materials helps understand structure-property relationships that may exist in such complicated structures.

\begin{acknowledgements}
A.G.\ acknowledges the hospitality of Los Alamos National Laboratory, where this project was initialized. 
She is also thankful to Dr.\ Lydie Louis, Dennis P.\ Trujillo, Dr.\ Ghanshyam Pilania and Dr.\ Geoffrey P.F.\ Wood for their helpful contributions 
to code development and discussions on implementation of various machine learning techniques. 
This work was supported by the U.S.\ DOE NNSA under Contract No.\ 89233218CNA000001 through the Rapid Response Program of Institute for Materials Science at LANL (A.G.), by the DOE BES ``Quantum Fluctuations in Narrow-Band System" Project (F.R),
and the NNSA Advanced Simulation and Computing Program (J.-X.Z.). It was supported in part through the Center for Integrated Nanotechnologies, a U.S.\ DOE Office of Basic Energy Sciences user facility in partnership with the 
LANL Institutional Computing Program for computational resources.
\end{acknowledgements}
%


\subsection{Contributions}
A.G.\ performed all computations and analysis and prepared all the figures. J.-X.Z.\ guided the project. 
F.R.\ provided insights on experimental reports and participated in the discussion of the results.
A.G.\ and J.-X.Z.\ wrote the paper with inputs from F.R.\ and S.N. 

\subsection{Additional Information}
The authors declare no competing interests.

\section{Supplementary Materials}

\subsection{Magnetic configurations}
\begin{figure}
\centering
\includegraphics[width=\columnwidth]{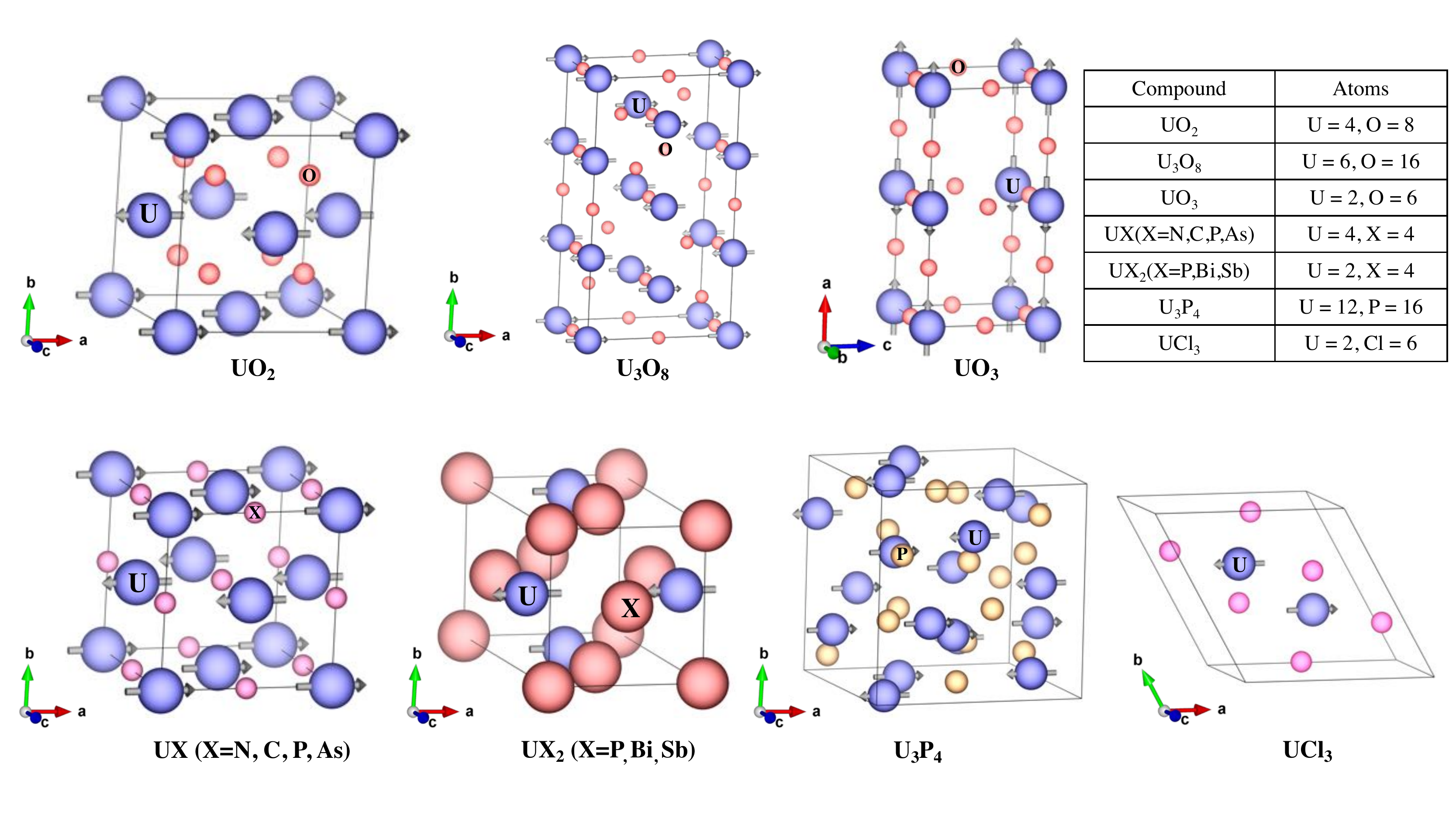}
\caption{(Color online) Structural Models (with AFM I configuration) of 12 compounds from Dataset I.}
\label{fig:mag_config}
\end{figure}
We note that these magnetic configurations are in accordance with that reported in literature.\cite{Burlet86,Troc66,Zhou2011,Wen2012,Lebegue2006} 
For U$_3$P$_4$, ferromagnetic configuration (FM) has a lower energy as compared to that in AFM I. 
This is also confirmed by additional DFT + Hubb U$_{\textit{eff}}$ (eV) computations where the energy difference between these two magnetic configurations is $\sim$3.26 eV.
The magnetic spins are assumed to be aligned along x axis in our results. \\
\begingroup
\squeezetable
\begin{table}[H]
\centering
\caption{Structural parameters and symmetry groups of 12 compounds from Dataset I as computed using DFT + Hubb U$_{\textit{eff}}$ (eV). All of these compounds have the same AFM I (self-consistently converged) configuration.}
\label{Magnetic_config}
\begin{ruledtabular}
\begin{tabular}{lcccc}
\\
                                             & Compound & lattice parameters (\AA)& Space Group & Hubb U$_{\textit{eff}}$ (eV)   \\             
\hline
\hline
(1) & UO$_2$ & a = b = c = 5.4682 & 225, Fm-3m & 0, 2, 4, 6                     \\
(2) & U$_3$O$_8$ & a = 6.7039, b = 11.9499, c = 4.1420 & 21, C222 & 0, 2, 4, 6       \\
(3) & UO$_3$ & a = 8.3299, b = 4.1649, c = 4.1649 & 221, Pm-3m & 0, 2, 4, 6 \\
(4) & UN & a = b = c = 4.4889 & 225, Fm-3m & 0, 2, 4, 6 \\
(5) & UC & a = b = c =  4.9597 & 225, Fm-3m & 0, 2, 4, 6 \\
(6) & UP &  a = b = c = 5.5869 & 225, Fm-3m & 0, 2, 4, 6 \\
(7) & UAs & a = b = c = 5.7767 & 225, Fm-3m & 0, 2, 4, 6 \\
(8) & UP$_2$ & a = b = 3.8099, c = 7.7639 & 129, P4/nmm & 0, 2, 4, 6 \\
(9) & UBi$_2$ & a = b = 4.4450, c = 8.9079 & 129, P4/nmm & 0, 2, 4, 6 \\
(10) & USb$_2$ & a = b = 4.2719, c = 8.7410 & 129, P4/nmm & 0, 2, 4, 6 \\
(11) & U$_3$P$_4$ & a = b = c = 8.2119 & 220, I-43d & 0, 2, 4, 6 \\
(12) & UCl$_3$ & a = b = 7.4429, c = 4.3210 & 176, P6$_3$/m & 0, 2, 4, 6 \\

\end{tabular}
\end{ruledtabular}
\label{table:moment}
\end{table}
\endgroup
\begingroup
\squeezetable
\begin{table}[H]
\centering
\caption{Differences in energy and moment sizes between spin-orientations chosen in-plane and out-of-plane.}
\label{Magnetic_config}
\begin{ruledtabular}
\begin{tabular}{lcccc}
\\
                                             & Compound & $\Delta$ E (eV) & $\Delta$ spinmom ($\mu$$_{\text{B}}$) & $\Delta$ orbitmom ($\mu$$_{\text{B}}$)   \\             
\hline
\hline
(1) & UO$_2$ & 0.032 & 0.001 & -0.009                   \\
(2) & U$_3$O$_8$ & 0.044 & 0.071 & -0.011      \\
(3) & UO$_3$ & 0.096 & 0 & 0 \\
(4) & UN & 0.007 & 0.015 & -0.037 \\
(5) & UC & 0.069 & 0.054 & -0.074 \\
(6) & UP & -0.003 & -0.090 & -0.068  \\
(7) & UAs & -0.002 & -0.006 & -0.02 \\
(8) & UP$_2$ & -0.206 & 0.008 & -0.036 \\
(9) & UBi$_2$ & 0.032 & 0.029 & 0.023 \\
(10) & USb$_2$ & 0.024 & 0.028 & 0.038 \\
(11) & U$_3$P$_4$ & 0.067 & -0.011 & 0.087  \\
(12) & UCl$_3$ & 0.279 & -0.009 & -0.011 \\

\end{tabular}
\end{ruledtabular}
\label{table:moment}
\end{table}
\endgroup

\subsection{Hyperparameters for ML models}
The optimized hyperparameters as obtained using a grid-search method for the regression algorithms are listed below.
The amount of penalization ($\alpha$)  is 0.01 for LASSO.
KRR uses a linear kernel with a regularization constant of 0.1
SVMR based models has used cost and epsilon parameters of 500 and 0.04 respectively.
RFR has used 60 decision trees and all the algorithms are also subjected to 10 cross-validations for each model.

\subsection{Statistical Bias}
\begin{figure}[H]
\centering
\includegraphics[width=\columnwidth]{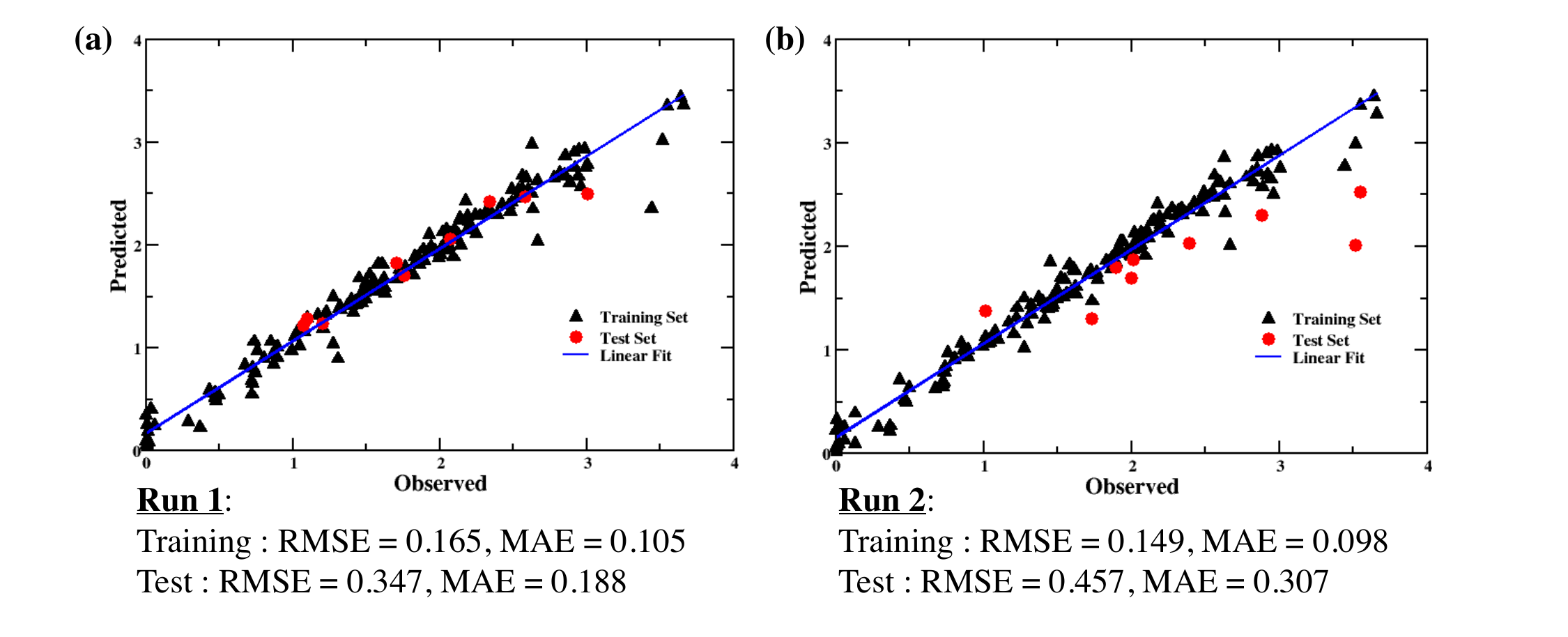}
\caption{(Color online) The root mean square and mean absolute errors vary between runs even when same model parameters are utilized due to statistical bias arising from small dataset size. }
\label{fig:stat_bias}
\end{figure}
\subsection{Correlation Maps for Dataset I for all features}
\begin{figure}[H]
\centering
\includegraphics[width=\columnwidth]{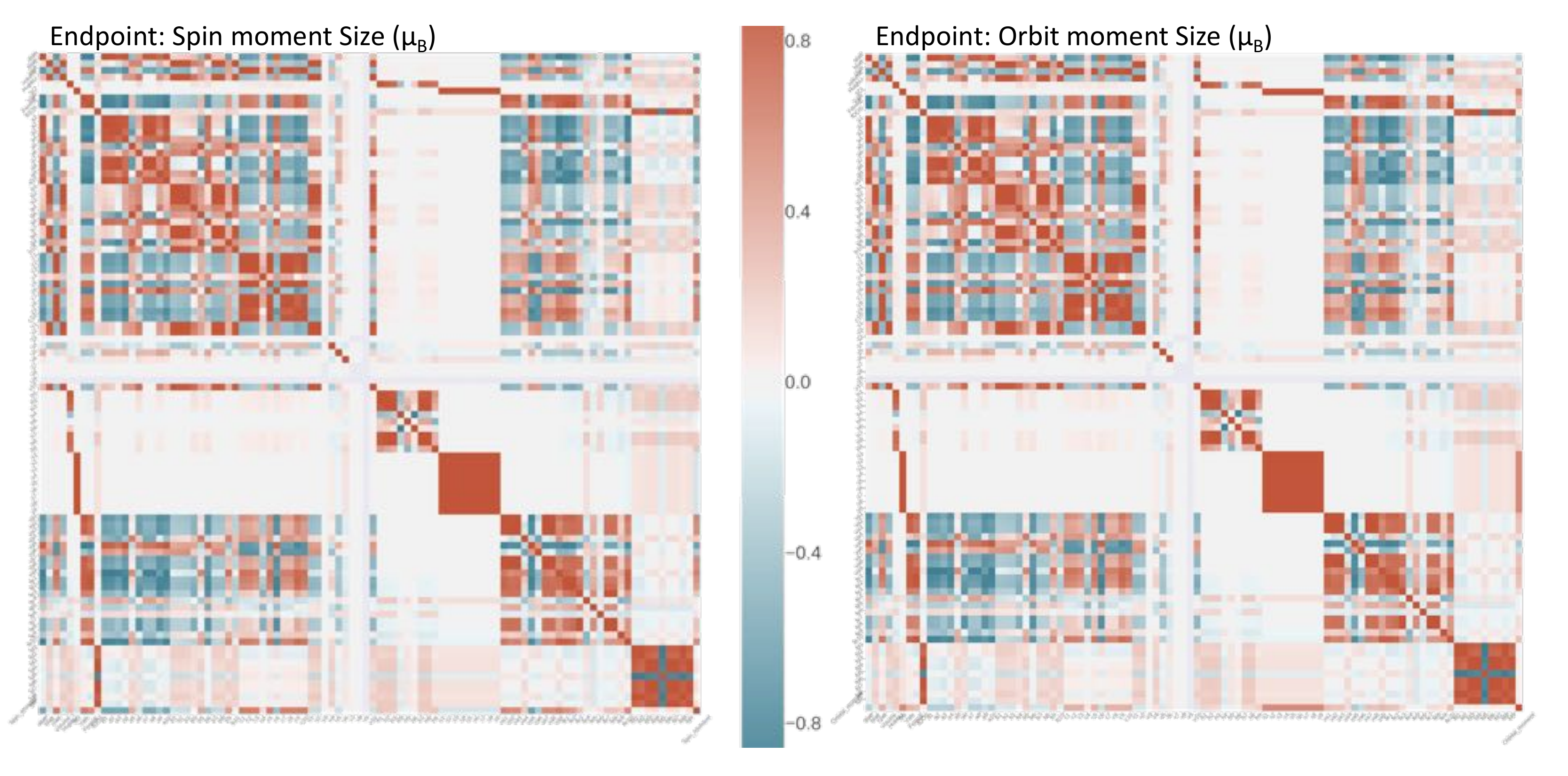}
\caption{(Color online) Correlation maps for combined set of primary and compound features.}
\label{fig:Corr_maps}
\end{figure}
\subsection{Decision Trees}
\begin{figure}[H]
\centering
\includegraphics[width=\columnwidth]{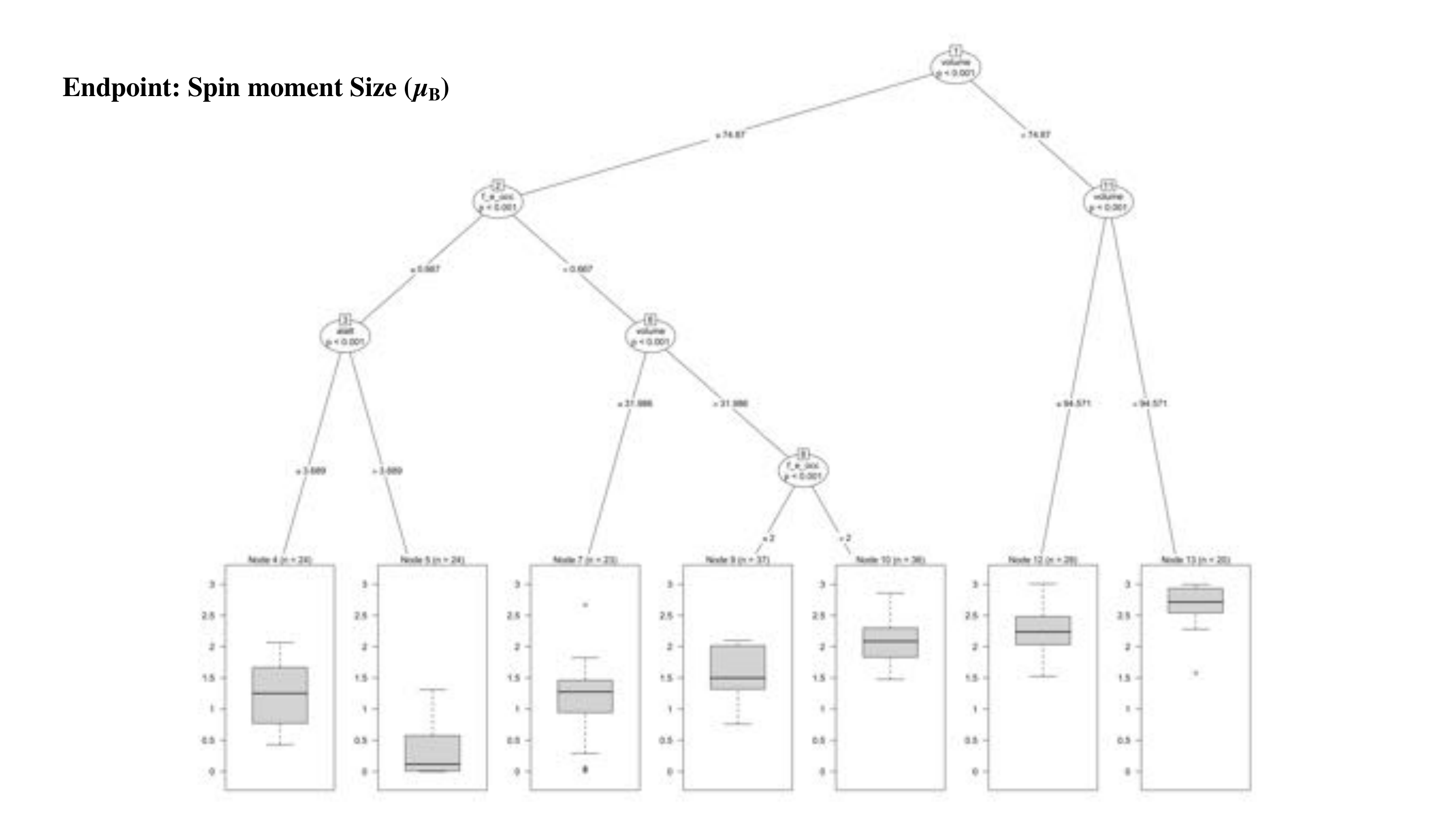}
\caption{(Color online) Conditional inference trees showing categorization of spin moment sizes based on primary features of all entries in Dataset I. Each node is described by the features used at the split. The split is chosen according to Bornferroni-corrected significance (p-value) which is also given in the diagrams. Vertical axes of the box plots signify the moment sizes whereas thick lines are median, boxes are quartiles, whiskers are non-outlier minimum and maximum and outliers are specified by dots. The number of response variables at each terminal node is also given.}
\label{fig:decision_trees}
\end{figure}
\begin{figure}[H]
\centering
\includegraphics[width=\columnwidth]{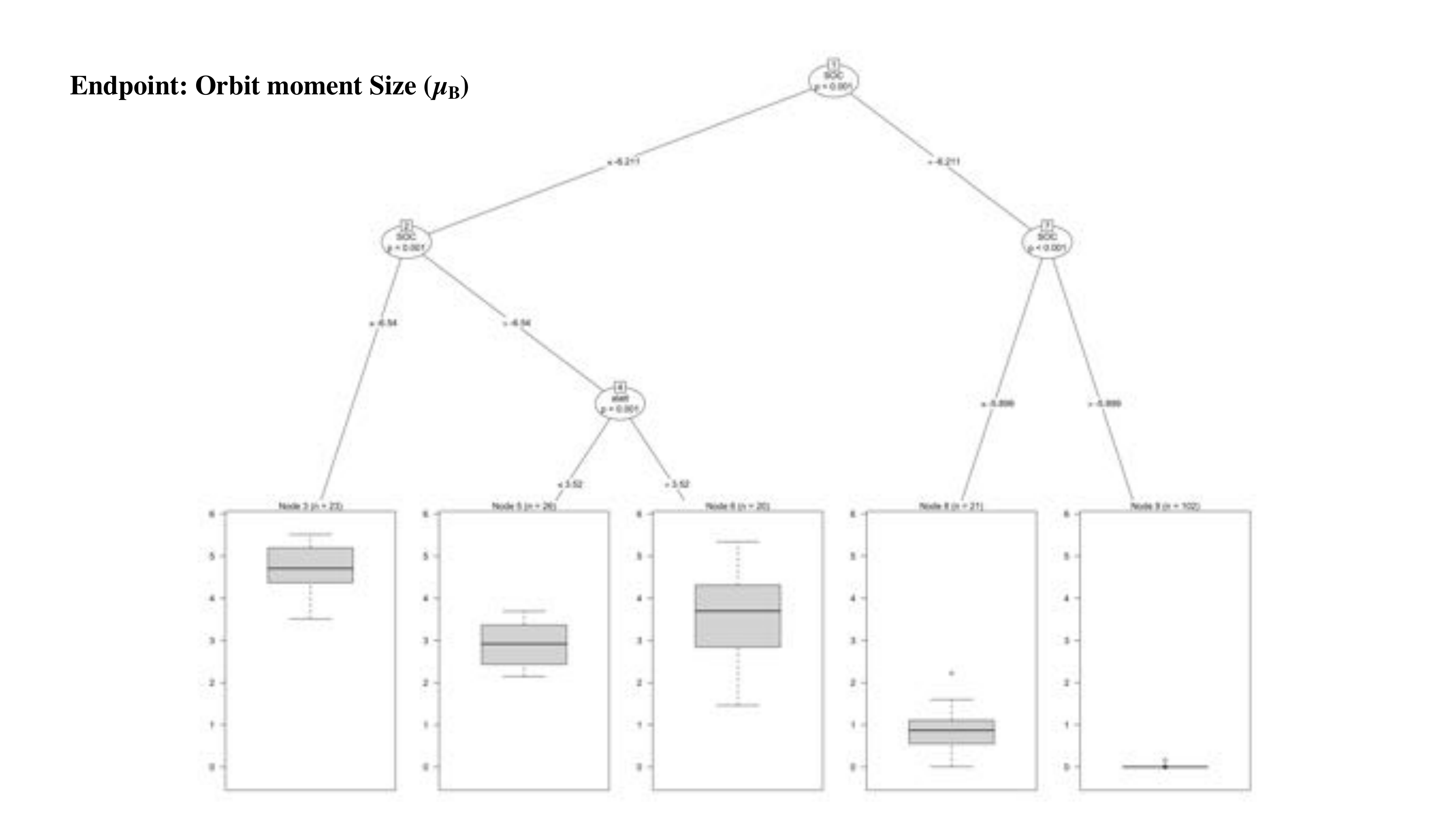}
\caption{(Color online) Conditional inference trees showing categorization of spin moment sizes based on primary features of all entries in Dataset I. Each node is described by the features used at the split. The split is chosen according to Bornferroni-corrected significance (p-value) which is also given in the diagrams. Vertical axes of the box plots signify the moment sizes whereas thick lines are median, boxes are quartiles, whiskers are non-outlier minimum and maximum and outliers are specified by dots. The number of response variables at each terminal node is also given.}
\label{fig:decision_trees}
\end{figure}
\subsection{ML models results}
\begin{figure}[H]
\centering
\includegraphics[width=\columnwidth]{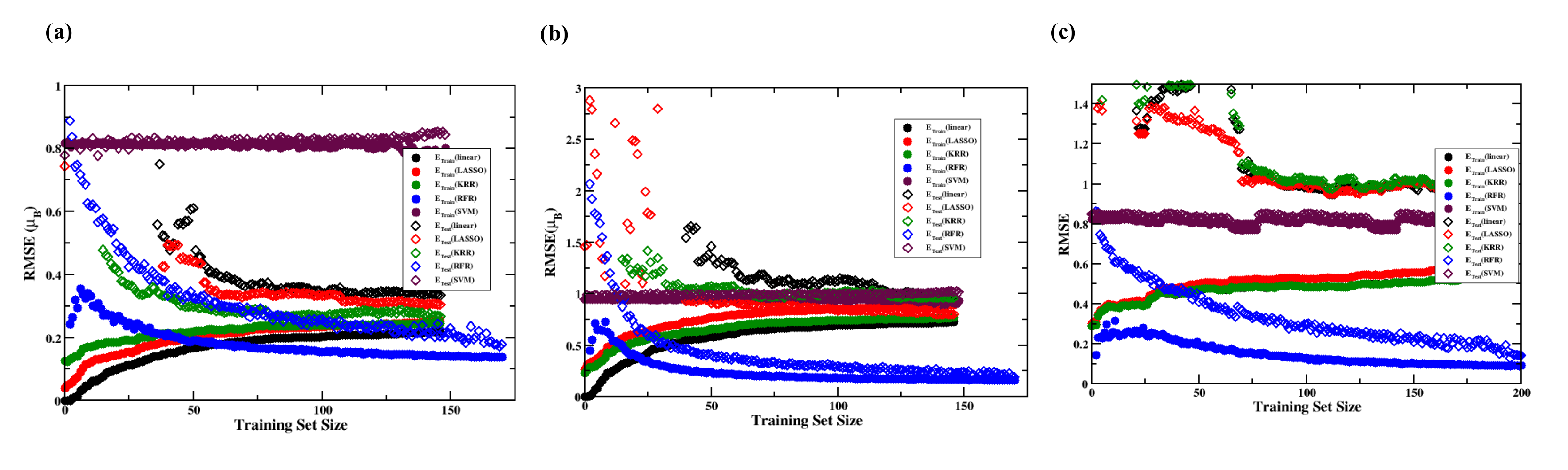}
\caption{(Color online) Comparative learning curves for machine learning models build using linear, LASSO, KRR, RFR and SVMR algorithms to predict end properties such as (a) spin moment size, (b) orbit moment size and (c) ordering respectively.}
\label{fig:LR_compare}
\end{figure}
\begingroup
\squeezetable
\begin{table}[H]
\centering
\caption{Test set errors and difference (overfitting) between the training and test set errors for the training set size of 50 and the largest training test size as obtained from learning curves for two machine learning models predicting ordering using RFR algorithms.}
\label{Magnetic_config}
\begin{ruledtabular}
\begin{tabular}{lccccc}
  & Endpoint & E$_{\text{Test}}$(\textit{50}) & E$_{\text{Test}}$$-$E$_{\text{Train}}$(\textit{50}) & E$_{\text{Test}}$(\textit{max}) & E$_{\text{Test}}$$-$E$_{\text{Train}}$(\textit{max}) \\             
\hline
 & Ordering & 0.52 & 0.32 & 0.20 & 0.12  \\
 & Ordering (using OFM) & 0.41 & 0.22 & 0.12 & 0.04  \\
\end{tabular}
\end{ruledtabular}
\label{table:errors_OFM}
\end{table}
\endgroup
\subsection{Hill Limit Comparison}
\begin{figure}[H]
\centering
\includegraphics[width=\columnwidth]{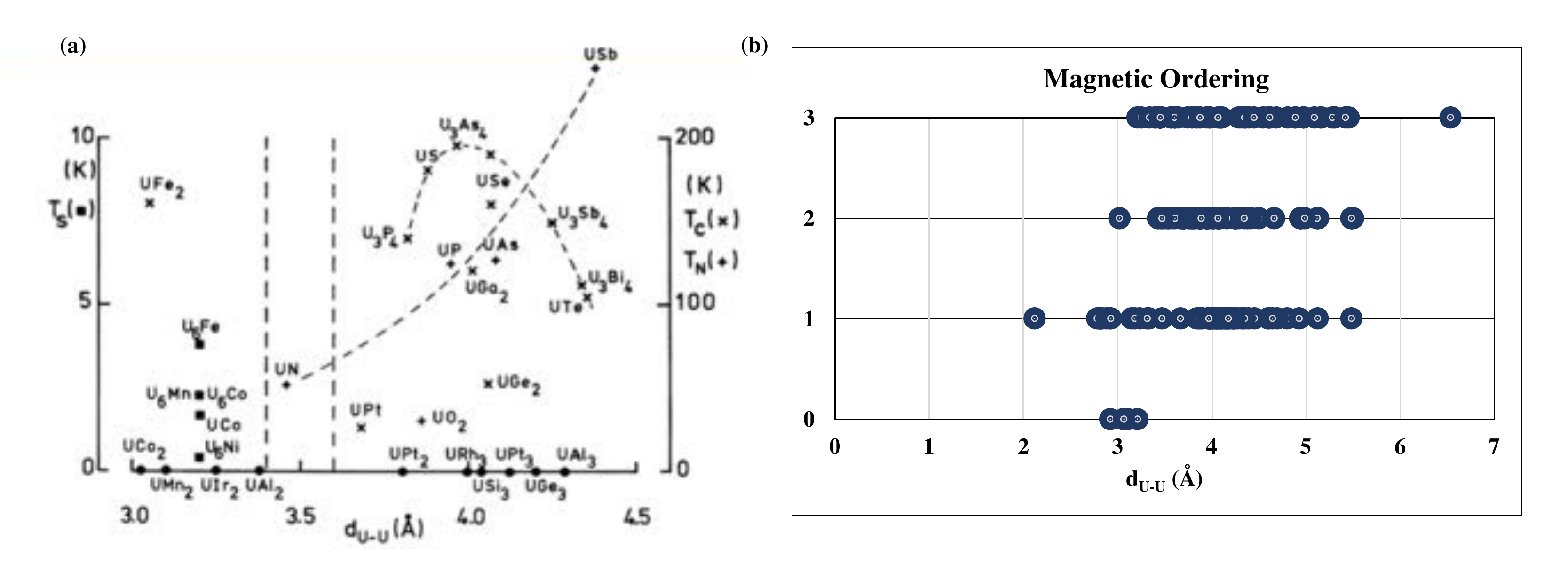}
\caption{(Color online) (a) Reference\cite{Hill70} Hill plot for a selected number of superconducting, paramagnetic, ferromagnetic and antiferromagnetic uranium compounds. (b) Average nearest neighbor distances for all entries in Dataset II plotted according to the corresponding observed ordering.}
\label{fig:hill_plot}
\end{figure}
\subsection{Information of Validation Set III}
External Validation Set III\cite{Leciejewicz,Brodsky,Gurtovoi,Hecker,Mentink,Griveau,Sechovsky,Troc2012,Gukasov2002,Purwanto1994,Samsel2008,Matsumoto2013,Ikeda2003,Bartha2015,Yarmolyuk1979} has compounds : 
U$_2$Co$_2$In,
U$_2$Co$_2$Sn,
U$_2$Co$_3$Si$_5$,
U$_2$Fe$_2$Sn,
U$_2$Ir$_2$Sn,
U$_2$Mo$_3$Ge$_4$,
U$_2$Mo$_3$Si$_4$,
U$_2$Nb$_3$Ge$_4$,
U$_2$Ni$_2$In,
U$_2$Ni$_2$Sn,
U$_2$Pd$_2$In,
U$_2$Pd$_2$Sn$_2$,
U$_2$Pt$_2$In,
U$_2$Pt$_2$Sn,
U$_2$PtC$_2$,
U$_2$Rh$_2$Sn,
U$_2$RhIn$_8$,
U$_2$Ru$_2$Sn,
U$_2$Ta$_3$Ge$_4$,
U$_2$Wi$_3$Si$_4$,
U$_3$Al$_2$Si$_3$,
U$_3$Au$_3$Sn$_4$,
U$_3$Co$_3$Sb$_4$,
U$_3$Cu$_3$Sb$_4$,
U$_3$Cu$_3$Sn$_4$,
U$_3$Cu$_4$Ge$_4$,
U$_3$Ir$_3$Sb$_4$,
U$_3$Ni$_3$Sb$_4$,
U$_3$Ni$_3$Sn$_4$,
U$_3$Ni$_4$Si$_4$,
U$_3$Pd$_3$Sb$_4$,
U$_3$Pt$_3$Sb$_4$,
U$_3$Pt$_3$Sn$_4$,
U$_3$Rh$_3$Sb$_4$,
U$_3$Rh$_4$Sn$_13$,
U$_3$Ru$_4$Al$_12$,
U$_4$Os$_7$Ge$_6$,
U$_4$Re$_7$Si$_6$,
U$_4$Ru$_7$Ge$_6$,
U$_4$Tc$_7$Ge$_6$,
U$_4$Tc$_7$Si$_6$,
UAsSe,
UAsTe,
UAu2A,l
UAu2In,
UAu2Sn,
UAuAl,
UAuGa,
UAuGe,
UAuSi,
UAuSn,
UCo$_2$Ge$_2$,
UCo$_2$P$_2$,
UCoAs$_2$,
UCoGa$_5$,
UCoGa,
UCoGa,
UCoGe,
UCoP$_2$,
UCoSi,
UCoSn,
UCoSn,
UCr$_2$Si$_2$,
UCrC$_2$,
UCu$_2$As$_2$,
UCu$_2$Ge$_2$,
UCu$_2$P$_2$,
UCu$_2$Si$_2$,
UCu$_2$Sn,
UCuBi$_2$,
UCuGa,
UCuGe,
UCuP$_2$,
UCuSb$_2$,
UCuSi,
UCuSn,
UFe$_2$Ge$_2$,
UFe$_2$Si$_2$,
UFeAl,
UFeAs$_2$,
UFeGa$_5$,
UFeGa,
UFeGe,
UFeSi,
UIr$_2$Ge$_2$,
UIr$_2$Si$_2$,
UIrAl,
UIrGe,
UIrSi$_3$,
UIrSi,
UIrSn,
UMn$_2$Ge$_2$,
UMn$_2$Si$_2$,
UNi$_1.6$As$_2$,
UNi$_2$Al$_3$,
UNi$_2$Ga,
UNi$_2$Ge$_2$,
UNi$_2$Si$_2$,
UNi$_2$Sn,
UNiAl,
UNiGa,
UNiGa$_3$,
UNiGa$_5$,
UNiGe,
UNiSb$_2$,
UNiSi,
UNiSn,
UOsGa$_5$,
UPd$_2$Al$_3$,
UPd$_2$Ga,
UPd$_2$Si$_2$,
UPd$_2$Sn,
UPdGa$_5$
UPdGe,
UPdIn,
UPdSb,
UPdSi,
UPdSn,
UPt$_2$Si$_2$,
UPt$_4$Au,
UPtAl,
UPtGa$_5$
UPtGe,
UPtIn,
UPtSi,
UPtSn,
URh$_2$Ge$_2$,
URhAl,
URhGa$_5$,
URhGe,
URhIn$_5$,
URhSi,
URhSn,
URu$_2$P$_{1.894}$,
URu$_2$Si$_2$,
URu$_2$Si$_2$,
URu$_4$B$_4$,
URuAl,
URuGa$_5$,
URuSb,
URuSn.

\end{document}